\documentclass[fleqn,usenatbib]{mnras}

\usepackage{mathptmx}

\usepackage[T1]{fontenc}
\usepackage{ae,aecompl}

\usepackage{graphicx}	
\usepackage{subcaption}
\usepackage{amsmath}	
\usepackage{amssymb}	

\usepackage[english]{babel}
\usepackage[utf8]{inputenc}
\usepackage{xspace}

\newcommand{\txs}{TXS\,0506$+$056\xspace}
\newcommand{\gray}{$\gamma$-ray}

\title[Lepto-hadronic single-zone models for the $\gamma$ and $\nu$ emission of \txs]{Lepto-hadronic single-zone models for the electromagnetic and neutrino emission of \txs}

\author[M. Cerruti et al.]{M. Cerruti$^{1, \star}$, A. Zech$^{2}$, C. Boisson$^{2}$, G. Emery$^{1}$, S. Inoue$^{3}$, and J.-P. Lenain$^{1}$\\ 
$^{1}$ Sorbonne Universit\'e, Universit\'e Paris Diderot, Sorbonne Paris Cit\'e, CNRS/IN2P3, Laboratoire de Physique Nucl\'eaire et de\\
Hautes Energies, LPNHE, 4 Place Jussieu, F-75252 Paris, France\\
$^{\star}$ matteo.cerruti@in2p3.fr\\
$^{2}$ LUTH, Observatoire de Paris, PSL Research University, CNRS, Universit\'e Paris Diderot, 5 Place Jules Janssen, 92190 Meudon, France\\
$^{3}$ RIKEN, Institute of Physical and Chemical Research, 2-1 Hirosawa, Wako, Saitama, 351-0198, Japan
}
\date{Accepted XXX. Received YYY; in original form ZZZ}

\pubyear{2018}

\begin{document}
\label{firstpage}
\pagerange{\pageref{firstpage}--\pageref{lastpage}}
\maketitle

\begin{abstract}
While active galactic nuclei with relativistic jets have long been prime candidates for the origin of extragalactic cosmic rays and neutrinos, the BL Lac object \txs is the first astrophysical source observed to be associated with some confidence ($\sim 3\sigma$) with a high-energy neutrino, IceCube-170922A, detected by the IceCube Observatory. The source was found to be active in high-energy \gray s with {\it Fermi}-LAT and in very-high-energy \gray s with the MAGIC telescopes. To consistently explain the observed neutrino and multi-wavelength electromagnetic emission of \txs, we investigate in detail single-zone models of lepto-hadronic emission, assuming co-spatial acceleration of electrons and protons in the jet, and synchrotron photons from the electrons as targets for photo-hadronic neutrino production. 
The parameter space concerning the physical conditions of the emission region and particle populations is extensively explored for scenarios where the \gray s are dominated by either 1) proton synchrotron emission or 2) synchrotron-self-Compton emission, with a subdominant but non-negligible contribution from photo-hadronic cascades in both cases. We find that the latter can be compatible with the neutrino observations, while the former is strongly disfavoured due to the insufficient neutrino production rate.
\end{abstract}



\section{Introduction}
\citet{TXS0506} recently reported the detection of a high-energy neutrino (IceCube-170922A, hereafter IC-170922A) with good angular resolution, for the first time coinciding spatially and temporally with a blazar in an elevated \gray\ flux state. A chance correlation is rejected at the 3$\sigma$ level. The blazar in question, \txs, is a BL Lac object, and a known emitter of 
high-energy \gray s \citep{fermicat}. During the multi-wavelength campaign triggered by the IceCube alert \citep{TXS0506, veritas0506, txs_magic,keivani18}, it was seen by {\it Fermi}-LAT to be in a high state that began in April 2017 and lasted several months, and was also discovered in very-high-energy (VHE; $E>100\,$GeV) \gray s with the MAGIC Cherenkov telescopes. A search for further neutrinos from \txs\ in the IceCube data found evidence at $3.5 \sigma$ for a neutrino flare in 2014-2015, and absence of additional neutrinos during the 2017 \gray\ flare \citep{icecube0506_paper2}. The redshift of the source was recently measured to be $z=0.337$ \citep{paiano18}. There is currently no estimate on the mass of the central black hole.

While the probability of 56.5$\%$ quoted by \citet{TXS0506} for this single neutrino to be truly astrophysical does not yet firmly establish blazars as sources of high-energy neutrinos, this detection represents the first direct observational indication for such a link. Earlier attempts to find correlations between high-energy neutrinos and blazars \citep[][]{Kadler16, lucarelli17} suffered from insufficient angular resolution of the neutrinos and the absence of  \gray\ signals well correlated in time, so that the significance of the association was marginal.

The simplest scenario that may explain correlated electromagnetic and neutrino emission in blazars is the one-zone, lepto-hadronic model \citep[see e.g.][]{mannheim93, aharonian00, mucke01}, where a magnetized compact region inside the relativistic jet carries a population of relativistic electrons and protons. Neutrinos are generated as part of the pion-decay chain in proton-photon interactions, while synchrotron-pair cascades of secondary particles and/or proton-synchrotron radiation are responsible for the high-energy part of the spectral energy distribution (SED), the low-energy part being usually ascribed to synchrotron radiation from primary electrons.

FSRQs are blazars more luminous than BL Lac objects and feature bright accretion disks that illuminate the surrounding medium, leading to intense photon fields that can serve as effective targets for pion-production. They are thus potentially capable of producing neutrinos with higher efficiency and luminosity compared to BL Lac objects \citep[see e.g.][]{murase14, rodrigues18}. However, it is difficult to interpret the electromagnetic emission from such luminous blazars with hadronic scenarios, without invoking excessive values for the power in accelerated protons \citep[see e.g.][]{zdz15}. On the other hand, for BL Lac objects, the only target photon field in the simplest one-zone model is provided by synchrotron radiation from primary electrons, implying a lower neutrino production efficiency. Nevertheless, the energetics requirements for BL Lac objects are lower \citep{leha-uhbl,cerruti17,zech17}, and several studies have proposed bright $\gamma$-loud BL Lac objects as potential sources of IceCube neutrinos \citep{padovani14,petropoulou15,righi17}.

In this Letter, we present an extensive exploration of the parameter space of the one-zone lepto-hadronic model applied to the high state of \txs, including solutions where the high-energy emission is dominated by either 1) proton-synchrotron radiation, or 2) synchrotron self-Compton (SSC) radiation from primary electrons (called ``mixed lepto-hadronic'' due to a non-negligible hadronic cascade contribution). We constrain the parameter space from the electromagnetic observations, and for each solution we compute the expected neutrino fluxes and IceCube event rates. Assuming that the association between \txs\ and IC-170922A is genuine, we then discuss the impact of the neutrino detection on blazar emission models. The dataset used in the paper is the one described in \citet{TXS0506}, restricting the optical data set to strictly simultaneous {\it Swift}-UVOT points.

\section{Numerical simulations}

The \textit{LeHa} code \citep{leha-uhbl} is used to simulate electromagnetic and neutrino emission from \txs. It has been developed to describe the stationary \gray emission from BL Lacertae objects, taking into account all relevant leptonic and hadronic radiative processes. 

The emitting region is a spherical plasmoid in a relativistic jet, parametrized by its radius $R$ and moving with Doppler factor $\delta$ with respect to the observer. The plasmoid is filled with a tangled, homogeneous magnetic field $B$, and a stationary population of leptons and hadrons, whose energy distributions are parametrized as broken power-law functions with exponential cut-offs. The hadronic part of the code simulates p-$\gamma$ interactions, and calculates the radiative output from all secondary particles. Photo-meson interactions are calculated using the Monte-Carlo code \textit{SOPHIA} \citep{sophia}, while Bethe-Heitler pair production is calculated following \citet{kelner08}. Photons from $\pi^0$ decay, secondary leptons from $\pi^\pm$ decay, and Bethe-Heitler pairs trigger synchrotron-supported pair-cascades in the emitting region. The low-energy photons that serve as targets for p-$\gamma$ interactions and $\gamma$-$\gamma$ pair-production are synchrotron photons from primary leptons and protons, and SSC photons. The energy distribution of secondary particles from hadronic processes is calculated by solving the corresponding differential equations including injection, and synchrotron, SSC, and adiabatic energy losses.

The fact that the parameter space of blazar hadronic models is much larger than that of leptonic models, together with the numerical challenges of correctly simulating hadronic interactions, make the search for best-fit solutions computationally prohibitive. The number of free parameters of the model is 15: 3 for the emitting region ($\delta$, $B$, and $R$), 12 for the primary population of leptons and protons (the 4 indices of the broken power-law distributions, $\alpha_{e/p , 1/2}$; the minimum, break, and maximum Lorentz factors $\gamma_{e/p,\textnormal{min/break/max}}$; and the normalizations $K_{e/p}$). We reduce the number of free parameters via the following physically motivated a priori assumptions:
\vspace{-0.25cm}
\begin{itemize}
\item assuming a common acceleration mechanism for the primary leptons and hadrons, we fix the spectral indices of the particle populations to be $\alpha_{e,1} = \alpha_{p,1}$ at injection;
\item the primary leptons lose energy mostly via synchrotron radiation, which imposes $\alpha_{e,2} = \alpha_{e,1} + 1$, and $\gamma_{e,\textnormal{break}}$ is determined by equating the synchrotron cooling time-scale $\tau_\textnormal{syn}$ with the adiabatic expansion time-scale $\tau_\textnormal{ad} = 2 R/c$;
\item the minimum Lorentz factor of the proton distribution $\gamma_{p,\textnormal{min}}$ is fixed to be 1, since its impact on the models is minor, as long as its value is not too large;
\item the same holds for the minimum Lorentz factor of the electron distribution $\gamma_{e,\textnormal{min}}$ which, however, cannot be arbitrarily low, in order not to overshoot archival radio data, and has been fixed to be $500$;
\item the maximum proton Lorentz factor $\gamma_{p,\textnormal{max}}$ is calculated by equating the acceleration time scale, parametrized as $\tau_\textnormal{acc} = \eta \frac{m_p c}{e B} \gamma_p$, and the shortest energy loss time scale for the protons. Constraints from the high-energy peak position on $\gamma_{p,\textnormal{max}}$ are thus translated into constraints on the parameter $\eta$ that controls $\tau_\textnormal{acc}$; as discussed in \citet{leha-uhbl}, this constraint fulfills the Hillas criterion;
\item the proton distribution is considered to be a simple power-law without a spectral break, as their energy losses are typically insignificant below $\gamma_{p,\textnormal{max}}$, ;
\item the observed variability time scale $\tau_\textnormal{obs}$, measured to be of the order of one day by MAGIC, is used to constrain the radius of the emission region as $R \leq \frac{\delta \,\, \tau_\textnormal{obs}}{1+z}$.
\end{itemize}
\vspace{-0.2cm}

The number of free parameters is thus reduced to eight: $\delta$, $B$, and $R$ 
for the emitting region, with $R$ limited via the variability timescale; and $K_e$, $K_p$, $\gamma_{e,\textnormal{max}}$, $\alpha_{e,1} = \alpha_{p,1}$, and $\eta$ for the particle distributions. $K_e$, and $\gamma_{e,\textnormal{max}}$ are adjusted to fit the low-energy SED component. The parameter space of the remaining six parameters is studied separately for the two scenarios under investigation. For each model we compute a posteriori the $\chi^2$ with respect to the data, identify the solution with the lowest $\chi^2$, and select all models which are characterized by a $\Delta \chi^2$ within a range of $\pm 1\sigma$. As is generally the case for one-zone blazar models, the available radio data cannot be reproduced with any of the solutions discussed here due to synchrotron self absorption, and are likely associated with more extended regions in the jet. 

The neutrino spectrum is extracted for each (anti-)~neutrino flavor and propagated to the observer frame. It is assumed that the total neutrino flux is distributed equally among the three flavours due to neutrino oscillations. The estimated muon neutrino flux is then convolved with the effective areas for the IceCube EHE trigger \citep{TXS0506} and the IceCube point-source search \citep[PS in the following,][]{icecube_ps_search} to estimate the detection rates.

\begin{figure}
\begin{center}
\begin{subfigure}{.39\textwidth}
  \centering
  \includegraphics[width=\linewidth]{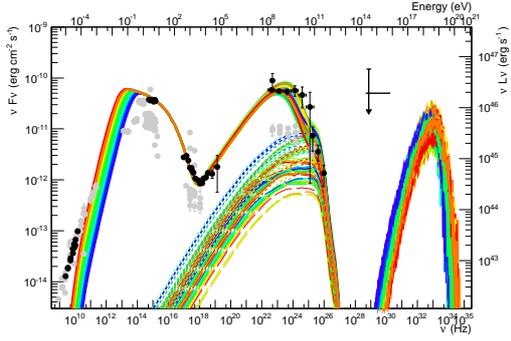}
  \caption{Proton synchrotron modeling of \txs}
  \label{plotpsyn}
\end{subfigure}%
\\
\begin{subfigure}{.39\textwidth}
  \centering
  \includegraphics[width=\linewidth]{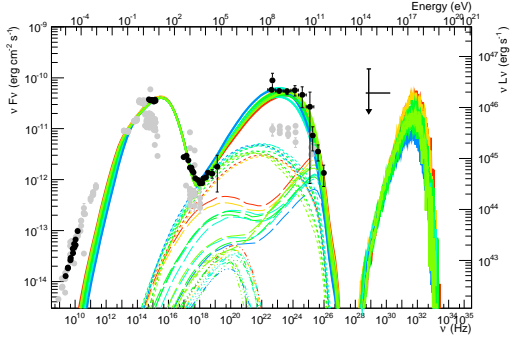}
  \caption{Lepto-hadronic modeling of \txs}
  \label{plotleha}
\end{subfigure}
\caption{Modeling of \txs\ for the proton synchrotron (\ref{plotpsyn}) and lepto-hadronic (\ref{plotleha}) scenarios. Black points are data from \citet{TXS0506}, while gray points are archival data. For each model, bold lines represent the total emission in photons (E $< 100$ TeV) and neutrinos (single flavour, E $> 100$ TeV); dashed lines the emission from pion cascades; dotted lines the emission from Bethe-Heitler cascades; dotted-dashed lines the proton synchrotron emission. Colours from red to blue represent increasing values of $R$. \vspace{-0.5cm}}
\label{sedplot}
\end{center}
\end{figure}

\begin{table}
    \centering
    \caption{Parameters used for the hadronic models }
   		\begin{tabular}{@{}l c c}
 		\hline
		 & Proton-synchrotron & Lepto-hadronic \\
 		\hline
 		 \noalign{\smallskip}
 		$\delta$ & $35-50$ & $30-50$  \\
 $R$ [10$^{16}$ cm] & $0.1-9.7$ & $0.2-1.5$\\
 $^\star \tau_\textnormal{obs}$ [days] & $0.01-1.0$& $0.02-0.3$ \\
 		\hline
 		 $B$ & $0.8-32$ & $0.13-0.65$  \\
 		$^\star u_B$ [erg cm$^{-3}$] & $0.02-0.16$ & $6.5\times10^{-4}-0.017$ \\
 		\hline
 		$\gamma_{e,\textnormal{min}} $& $500$ & $500$  \\
 		$\gamma_{e,\textnormal{break}} $& $=\gamma_{e,\textnormal{min}}$ & $=\gamma_{e,\textnormal{max}}$     \\
 		$\gamma_{e,\textnormal{max}}\ [10^4]$& $0.6-1.0$ & $0.8-1.7$\\
 		$\alpha_{e,1}=\alpha_{p,1}$ & $2.0$  & $2.0$ \\
 		$\alpha_{e,2}=\alpha_{p,2}$ & $3.0$  & $3.0$ \\
 		$K_e$ [cm$^{-3}$] & $6.3-9.1\times10^{3}$   &$9.5\times10^{3}-2.6\times10^{5}$  \\
 		$^\star u_e$ [10$^{-5}\,$erg$\,$cm$^{-3}$] & $0.4-15.1$  & $2.2\times10^{3}-43\times10^{3}$ \\
 		\hline
 		$\gamma_{p,\textnormal{min}}$& 1 & 1  \\
 		$\gamma_{p,\textnormal{break}} [10^9]$&  $=\gamma_{p,\textnormal{max}}$ & $=\gamma_{p,\textnormal{max}}$  \\
 		$\gamma_{p,\textnormal{max}} [10^9]$& $0.4-2.5$ & $0.06-0.2$\\
         $\eta$ & $20-50$ & $10$ \\
 		$K_p$ [cm$^{-3}$] & $10.4-2.0\times10^{4}$ & $3.5\times10^{3}-6.6\times10^{4}$ \\
 		$^\star u_p$ [erg cm$^{-3}$] &$0.7-45$ &  $100-1400$  \\
 		\hline
    	$^\star u_p/u_B $ & $1.0-89$  & $3.9\times10^{4}-79\times10^{4}$ \\
 		$^\star L$ [10$^{46}$ erg s$^{-1}$] & $0.8-170$ & $35-350$  \\
		\hline
        $^\star \nu_{\textnormal{EHE}}$ [yr$^{-1}$] & $5.7\times10^{-3}-0.16$ & $0.11-3.0$\\
        $^\star \nu_{\textnormal{EHE}, (0.183-4.3)\,\textnormal{PeV}}$ [yr$^{-1}$] &$2.4\times10^{-5}-1.7\times10^{-3}$ & $0.008-0.11$ \\
        $^\star \nu_{\textnormal{PS}}$ [yr$^{-1}$] & $0.011-0.32$ & $0.3-6.9$\\
 		\hline
 		\end{tabular}
 	 \label{table1a}
 	 \newline
 		 The luminosity of the emitting region has been calculated as $L=2 \pi R^2c\Gamma_\textnormal{bulk}^2(u_B+u_e+u_p)$, where $\Gamma_\textnormal{bulk}=\delta/2$, and $u_B$, $u_e$, and $u_p$, the energy densities of the magnetic field, the electrons, and the protons, respectively. The quantities flagged with a star ($^\star$) are derived quantities, and not model parameters. The full set of parameters is available as online material.
 		\end{table}

\subsection{Proton synchrotron solutions}
In a first approach, we ascribe the high-energy peak of the SED to proton-synchrotron emission, with sub-dominant contributions from synchrotron-pair cascades. As $\gamma_{p,\textnormal{max}}$ is defined by equating the acceleration and cooling time-scales, there exists a maximum proton-synchrotron peak frequency $\nu_{\textnormal{max}}$, for a given choice of $\delta$ and  $\eta$. We initially set $\eta = 10$ as in \citet{leha-uhbl}. For the maximum allowed value of $\nu_{\textnormal{max}}$, the energy of the proton-synchrotron peak is too high compared with the data. Lowering $\nu_{\textnormal{peak},p}$ leads to a denser emission region with a larger contribution from cascades. Adjusting the peak energy to agree with the data, without over-predicting the VHE and hard X-ray emission due to the cascade component, requires an increase in the value of $\eta$, i.e. a lower efficiency of the acceleration process.

The transition between the low-energy and high-energy component in the SED is well constrained by the combination of the {\it Swift}/XRT and {\it NuSTAR} data. A large contribution of the cascade component to the {\it NuSTAR} band is disfavoured, as it would invariably overproduce the VHE emission due to its broad spectral coverage. The only alternative is to adjust the spectral slope of the primary particle spectrum so that the proton-synchrotron component dominates the SED from the {\it NuSTAR} band up to the high-energy peak. The index of the primary particle distributions is thus fixed to a value of $2.0$.

In this scenario, the electrons are in the fast-cooling regime. Given the constraint on the co-acceleration of leptons and hadrons, the large value of the spectral index for particle injection leads to strong electron-synchrotron flux in the optical and infrared range.

These various constraints imply a well defined region in the parameter space. We scanned the following range of parameters: $\delta \in [20-50]$, with seven bins linearly spaced; $R \in [10^{15} \textnormal{cm}-R_\textnormal{max}]$, with ten bins logarithmically spaced; $\nu_{\textnormal{peak},p} \in [\nu_\textnormal{max} / 1000, \nu_\textnormal{max}]$ with ten bins logarithmically spaced;  $\eta \in [10, 50]$, with five bins linearly spaced; and $K_p \in [K^\star / 3, 3 K^\star]$, with five bins logarithmically spaced, where $K^\star$ corresponds to the proton density such that the peak of the proton synchrotron component is at the level of the {\it Fermi}-LAT data. In total we produce 17500 different models, and we use the $\chi^2$ to identify $1\sigma$ contours for the different parameters, as listed in Table~\ref{table1a}.

Allowing for values of $\delta$ up to 50, solutions within the $1 \sigma$ confidence region were found for $B = 0.8 - 32$\,G and $R = 10^{15-17}$\,cm. No satisfactory solutions were found with $\delta < 35$. 

The jet power required for this scenario can vary from $8 \times 10^{45}$ to $1.7 \times 10^{48}$\,erg s$^{-1}$, increasing with $R$. The solutions with the lowest power are thus well below the Eddington luminosity of a $10^9$ M$_\odot$ black hole. The ratio of energy densities in protons to magnetic fields is $u_p/u_B = 1 - 89$, being closer to equipartition for lower total power.

The neutrino spectra for the proton-synchrotron solutions are shown in Fig.~\ref{plotpsyn}. Their shapes are narrow, typically peaking above $10^{18}$\,eV. The expected detection rates vary widely among the different solutions, yielding an EHE muon neutrino rate between $5.7 \times 10^{-3}$ and $0.16$ yr$^{-1}$.
When restricting the estimate to the reported 90\% uncertainty on the energy of IC-170922A ($183$ TeV-$4.3$ PeV), the rate drops significantly to between $2.4 \times 10^{-5}$ and $1.7 \times 10^{-3}$. For the highest neutrino rate, the Poisson probability of detecting one neutrino with an energy compatible with that measured by IceCube during the 6-month high state of the source is 0.085\%. The probability of detecting no events outside of this energy range over the same period is 82\%. The neutrino rates for the IceCube PS search are similarly low (0.011-0.32 yr$^{-1}$). Given the very hard neutrino spectra, the rates computed above $4.3$ PeV are identical, and the probability of non-detection is between 85\% and 99\%.

Similar results were independently obtained by \citet{keivani18}, although their solutions are in a different part of the parameter space ($B = 85$ G, $\delta = 5-15$), mainly due to their adoption of a different data set, notably in the optical-UV, as well as their omission of VHE data as additional constraints. \citet{gao18} also show that a proton synchrotron scenario yields a very low neutrino flux.

\subsection{Mixed lepto-hadronic solutions}
In an alternative approach, we interpret the high-energy emission as a combination of SSC and synchrotron radiation from the hadronic cascades, probing a very different region in parameter space, with notably smaller values of $B$. In this mixed lepto-hadronic scenario, the proton-synchrotron peak is hidden below the other components, and protons reach lower energies than in the proton-synchrotron scenario. The higher proton density in this scenario is expected to lead to a higher neutrino flux peaking at lower energies.

It is possible to interpret the broadband emission with a standard SSC model. When assuming a minimum electron Lorentz factor $\gamma_{e,\textnormal{min}}$ of a few hundred, the {\it NuSTAR} data can no longer be ascribed to SSC emission, but are accounted for by the cascade component. These data thus serve as upper limits to the hadronic component.

The index of the primary particle distributions is fixed to 2.0 as for the proton-synchrotron solutions, but in this parameter range the lepton population is not completely cooled. To ease the study of the parameter space, we limit our solutions to the case where  $\gamma_{e,\textnormal{break}} \geq \gamma_{e,\textnormal{max}}$. Given that acceptable solutions are found for $\eta = 10$, we also do not explore different values for this parameter. We scanned the following parameter space: $\delta \in [20, 50]$, with seven bins linearly spaced; $R \in [10^{15}\,\textnormal{cm}, R_\textnormal{max}]$, with ten bins logarithmically spaced; $\nu_{\textnormal{peak},p} \in [1.5\times10^{-8}\nu_\textnormal{max}, 1.5\times10^{-5}\nu_\textnormal{max}] $ with ten bins logarithmically spaced; $K_p \in [K^\star / 3, 3 K^\star]$, with five bins logarithmically spaced. In total we produce 3500 different lepto-hadronic models. Good solutions are found in a small region of parameter space with $B = 0.1 - 0.7$\,G and $R = 2 \times 10^{15} - 1.5 \times 10^{16}$\,cm for $\delta = 30 - 50$.

Solutions with lower $\delta$, down to $\delta = 20$, can be found when allowing for a detectable cooling break in the primary electron spectrum. Such solutions also provide a better representation of the optical data. However, for the automated parameter scan to be applicable to the mixed lepto-hadronic scenario, we have restricted this study to cases of uncooled electron distributions, while verifying that there is no significant impact on the modelling of the high-energy spectrum and the resulting ranges of jet power and neutrino fluxes. In all solutions, the SSC emission is largely dominating the high-energy peak, while the lower but flatter cascade emission spectrum is responsible for most of the hard X-rays and VHE $\gamma$-rays. The jet power is smallest for intermediate $\delta$, large $B$ and small $R$. The minimum value is $3.5\times10^{47}$ erg s$^{-1}$, about $40$ times larger than the minimum found for proton-synchrotron solutions. A denser proton population is needed to compensate for the weaker $B$. Values of $u_p$/$u_B \simeq 10^4 - 10^6$ are inferred, indicating the energetics to be far out of equipartition.

The neutrino spectra in the lepto-hadronic solutions are shown in Fig.~\ref{plotleha}. The flux level is higher than in proton-synchrotron scenarios, and the spectrum peaks at lower energies, typically below $10^{18}$\,eV. The estimated neutrino detection rate is between $0.1$ and $3.0$ per year for the parameter space we studied, but it should be noted that there is no actual lower limit to this rate in the lepto-hadronic scenario, if one allows for a sub-dominant contribution of the cascade component to the hard X-ray band.

The highest neutrino rates correspond to solutions with the highest total kinetic energy in protons ($\propto u_p R^3$) and intermediate jet power. Detection rates of more than $0.5$ yr$^{-1}$ can be attained even with a jet power close to the minimum value.
When restricting the estimate to the ($0.183-4.3$) PeV band, the detection rate is $0.008-0.11$ yr$^{-1}$. For the solution that provides the highest neutrino rate, the Poisson probability for detecting one $\nu_\mu$ with the energy measured by IceCube during the high-state is 5.2\%, while the probability for not detecting any events outside of the reconstructed energy interval is 5.5\%. The neutrino rates obtained with the PS effective area are much higher ($0.3-6.9$ yr$^{-1}$). Even when limiting the energy band to energies higher than 4.3 PeV, the expected rates remain $0.2-6.4$ yr$^{-1}$, indicating that these solutions predict multi-PeV neutrinos in addition to IceCube 170922A. The Poisson probability of detecting no neutrinos outside the energy range are  between 4\% and 88\%: the solutions with the highest EHE rates may thus face difficulties in explaining why only one neutrino was seen with IceCube. It is important to recall here that this conclusion does depend on the assumption that $\gamma_{p, \max}$ is linked to the acceleration timescale with $\eta = 10$. By relaxing this hypothesis, the neutrino spectra can peak at lower energies, lowering the expected rates.

\citet{gao18} also presented single-zone lepto-hadronic solutions, finding neutrino rates lower than our values. This may be due to different approaches in constraining the parameter space: while we kept $\gamma_{p, \max}$ as a free parameter (although linked to $B$ and $R$ via the balance of acceleration and cooling time-scales), they searched for solutions for two fixed values of $\gamma_{p, \max} = 4.8 \times10^{6}$ and $7.5 \times10^{10}$, which are respectively much lower and higher than in our solutions.

\section{Discussion}

The probability of detecting a muon neutrino with energy inside the reported 90\% confidence interval of IC-170922A during the six-month high state of the \txs is sufficiently high for the lepto-hadronic scenario, but only marginal ($8.5\times10^{-4}$ at most) for the proton-synchrotron case. It should be noted, however, that the uncertainty on the energy of IC-170922A is large, and probabilities for detecting a neutrino increase rapidly when allowing for higher upper limits on the neutrino energy, due to the steeply increasing neutrino fluxes with energy expected in both scenarios. 

The available data set is very constraining for one-zone models, thanks to the good multi-wavelength coverage from the optical to the VHE range. While proton-synchrotron solutions are subject to degeneracy between $R$ and $B$ as found in earlier studies, the mixed lepto-hadronic solutions cover relatively small regions in the $B$-$R$ parameter space that would indicate an emitting region of typical extension $10^{16}$\,cm and a location at sub-parsec distance from the central engine for typical jet opening angles. 
As is usually found for (lepto-)hadronic models, the required jet power is relatively high and largely dominated by that in protons. Solutions in the proton-synchrotron scenario are generally less demanding in this respect, and one can find parameter sets likely corresponding to sub-Eddington luminosity.
It is important to underline that our solutions are characterized by $\alpha_{p,1} = 2.0$ and $\gamma_{p, \min} = 1$, and are thus conservative in terms of the total power in hadrons. A lower proton luminosity can be achieved if $\gamma_{p, \min} \geq 1$ or $\alpha_{p,1} \leq 2.0$. It can also be realized if the target photons for p-$\gamma$ interactions originate outside the jet \citep[e.g.][]{txs_magic, keivani18}.

The two scenarios should in principle be distinguishable with future variability studies of this source. While the lepto-hadronic scenario predicts a strong correlation between the low-energy and high-energy spectral bumps, as in any SSC scenario, the proton-synchrotron scenario would imply delays between variations in the two components due to the different acceleration and cooling time scales. A delay is also expected between the hard X-ray component and the high-energy peak flux in the lepto-hadronic solutions. In both scenarios, the time-averaged SED during the high state of the source is well reproduced by the model, while a rapid flux increase over a few nights, potentially seen in the VHE band, would require time-dependent modelling. The stationary solutions presented here are however consistent with a variability time scale of one day.

In the proton-synchrotron scenario, protons can reach a maximum Lorentz factor of $10^9$. Accounting for Doppler boosting, this can be close to the highest observed energies of ultra-high-energy cosmic rays, if the protons can escape the source without further energy losses. Maximum proton energies are lower by a factor of ten in the lepto-hadronic scenario, but the proton density is much higher. 

\section{Conclusions}
After introducing a few simplifying constraints based on general physical considerations, we have explored extensively the parameter space of the lepto-hadronic one-zone model for the SED of the 2017 high state of TXS\,0506+056. Good solutions can be found with the proton-synchrotron and mixed lepto-hadronic scenarios in restricted parameter regions. While the proton-synchrotron solutions are disfavoured if IC-179022A has its origin in the source, lepto-hadronic solutions can account for this event, while being more demanding in terms of the jet power. In addition, they are constrained by the nondetection so far of neutrinos with energies higher than that of IC-170922A with the IceCube point-like search algorithm.
If a second neutrino coincident with a \gray\ flare is detected from this source in the future, this would favour the lepto-hadronic scenario, while the absence of any future neutrino detections could be used to put constraints on the acceptable parameter space for both scenarios.

\section*{Acknowledgements}
\small{We gratefully acknowledge CC-IN2P3 (\href{https://cc.in2p3.fr}{cc.in2p3.fr}) for providing a significant amount of the computing resources and services needed for this work. Part of this work is based on archival data, software or online services provided by the Space Science Data Center - ASI. This work is supported by JSPS KAKENHI Grant Number JP17K05460 for SI. We thank the anonymous referee for his/her constructive comments which improved the quality of the paper.}




\bibliographystyle{mnras}
\bibliography{txs0506} 


\bsp	
\label{lastpage}
\end{document}